\documentclass{aa}

\usepackage{graphicx}
\usepackage{txfonts}
\usepackage{natbib}
\bibpunct{(}{)}{;}{a}{}{,} 

\begin{document}

\title{The origin of redshift asymmetries:\\How $\Lambda$CDM explains anomalous
redshift}

\author{S.-M. Niemi\inst{1,}\inst{2}
          \and
          M. Valtonen\inst{1}}

\institute{University of Turku, Department of Physics and Astronomy, Tuorla
		  Observatory, V\"ais\"al\"antie $20$, Piikki\"o, Finland\\
          \email{saniem@utu.fi}
         \and
          Nordic Optical Telescope, Apartado $474$, E$-38700$ Santa Cruz de
          La Palma, Santa Cruz de Tenerife, Spain\\}

\date{Released 2008}

\abstract
{}
{Several authors have found a statistically significant excess of galaxies with
higher redshifts relative to the group centre, so-called discordant redshifts, in
particular in groups where the brightest galaxy, identified in apparent
magnitudes, is a spiral. Our aim is to explain the observed redshift excess.}
{We use a semi-analytical galaxy catalogue constructed from the Millennium
Simulation to study redshift asymmetries in spiral-dominated groups in the
$\Lambda$cold dark matter ($\Lambda$CDM) cosmology. We create two mock catalogues
of galaxy groups with the Friends-of-Friends percolation algorithm to carry out
this study.}
{We show that discordant redshifts in small galaxy groups arise when these groups
are gravitationally unbound and the dominant galaxy of the group is
misidentified. About one quarter of all groups in our mock catalogues belong to
this category. The redshift excess is especially significant when the apparently
brightest galaxy can be identified as a spiral, in full agreement with
observations. On the other hand, the groups that are gravitationally bound do not
show a significant redshift asymmetry. When the dominant members of groups in
mock catalogues are identified by using the absolute B-band magnitudes, our
results show a small blueshift excess. This result is due to the magnitude
limited observations that miss the faint background galaxies in groups.}
{When the group centre is not correctly identified it may cause the major part of
the observed redshift excess. If the group is also gravitationally unbound, the
level of the redshift excess becomes as high as in observations. There is no need
to introduce any ``anomalous'' redshift mechanism to explain the observed
redshift excess. Further, as the Friends-of-Friends percolation algorithm picks
out the expanding parts of groups, in addition to the gravitationally bound group
cores, group catalogues constructed in this way cannot be used as if the groups
are purely bound systems.}



\maketitle

\section[]{Introduction}\label{s1}
Groups of galaxies contain a large fraction of all galaxies in the Universe
\citep{1950MeLu2.128....1H,Humason:1956p593,Huchra:1982p1,Geller:1983p576,Nolthenius:1987p357,Ramella:2002p514}.
These density enhancements in the sky and in the redshift space are important
cosmological indicators of the distribution of matter in the Universe, and may
provide important clues for galaxy formation. Groups of galaxies are, in general,
divided into a large number of different classes, for example, loose groups
\citep[e.g.][]{Ramella:1995p562,Tucker:2000p453,Einasto:2003p590}, poor groups
\citep[e.g.][]{Zabludoff:1998p578,Mahdavi:1999p581}, compact groups
\citep[e.g.][]{1973Afz.....9..495S,Hickson:1982p571,Hickson:1989p573,Focardi:2002p582}
and fossil groups
\citep[e.g.][]{1994Natur.369..462P,Jones:2003p594,DOnghia:2005p420,Santos:2007p217}.
However, from the observational point of view, groups of galaxies and their
member galaxies are not extremely well defined.

In recent years a number of grouping algorithms have been developed and applied
\citep[e.g.][]{Turner:1976p356,Materne:1978p353,Huchra:1982p1,2004MNRAS.349..425B,Goto:2002p585,Kim:2002p498,Bahcall:2003p227,Koester:2007p588,Yang:2007p506}
to identify real groups. Despite the vast number of grouping algorithms, the
Friends-of-Friends (FoF; \citealt{Huchra:1982p1}, hereafter HG82) percolation
algorithm remains the most frequently applied. The FoF algorithm or slightly
modified versions of it are widely used even for modern day galaxy surveys.
Several authors have presented group and cluster catalogues that applied the FoF
algorithm based on the SDSS (Sloan Digital Sky Survey) data
\citep[e.g.][]{Merchan:2005p495,Berlind:2006p484,Tago:2008p508} and on the 2dF (The
Two Degree Field) data \citep[e.g.][]{Eke:2004p509,Tago:2006p490}.

Few studies \citep[see][and references therein]{Niemi:2007p344} have argued that
grouping algorithms may not always return true groups; a significant number
of groups can be spurious and contain interlopers. \citet{Niemi:2007p344}
(hereafter Paper I) have shown that the FoF algorithm produces a significant
fraction of groupings which are not gravitationally bound systems, but merely
groups in a visual sense. This can introduce various errors when these groups are
studied in detail and treated as gravitationally bound structures.

This may even be true for compact groups of galaxies. In this case, and also
in general, extended X-ray emission can be used to distinguish real,
gravitationally bound groups from spurious ones. The analysis of X-ray data
suggests that errors are rather common \citep[][and references
therein]{1995ApJ...444L..61O}. Unfortunately, groups with low mass and
a spiral-dominated main galaxy in general do not show any extended X-ray emission.
Therefore this method is not useful for small and loose groups of galaxies.

An excess of higher redshift galaxies was discovered by \citet[][]{1970Natur.225.1033A, Arp:1982p577} and it was studied in detail by
\citet{1971Natur.234..534J}. Since then many authors have found a statistically
significant excess of high redshift companions relative to the group centre.
\citet{Bottinelli:1973p352} extended the study of \citet{1970Natur.225.1033A} to
nearby groups of galaxies in which the magnitude difference between the companion
and the main galaxy was greater than $0.4$ mag. \citet{Sulentic:1984p341} found a
statistically significant excess of positive redshifts while studying
spiral-dominated (i.e. the central galaxy is a spiral galaxy) groups in the
catalogue of galaxy groups by HG82, and derived the redshift excess $Z = 0.21$
for spiral-dominated groups while the E/S0 dominated (i.e. the central galaxy is
an E/S0 galaxy) groups showed a blueshift excess $Z = -0.13$. \citet{Girardi:1992p339} found discordant redshifts while studying nearby small groups identified by \citet{1988ngc..book.....T} in
the Nearby Galaxy Catalogue. However, the conventional theory holds that the
distribution of redshift differentials for galaxies moving under the
gravitational potential of a group should be evenly distributed. Even systematic
radial motions within a group would be expected to produce redshift differentials
that are evenly distributed.

Multiple theories have been suggested to explain the observed redshift
excess. \citet{Sulentic:1984p341} listed some possible origins for the observed
redshift excess. \citet{Byrd:1985p342} and \citet{Valtonen:1986p340} argued that
this positive excess is mainly due to the unbound expanding members and the fact
that the dominant members of these groups are sometimes misidentified.
\citet{Girardi:1992p339} argued that the positive excess may be explained if
groups are still collapsing and contain dust in the intragroup medium.
\citet{Hickson:1988p358} ran Monte Carlo simulations and concluded that the
random projection can explain discordant redshifts. \citet{Iovino:1997p338} found
that projection effects alone can account for the high incidence of discordant
redshifts. However, studies by \citet{Hickson:1988p358} and
\citet{Iovino:1997p338} dealt only with Hickson's compact groups of galaxies.
\citet{Tully:1987p355} analyzed his catalogue of nearby groups of galaxies and
did not find evidence of redshift asymmetries in galaxy groups. However, unlike
in earlier work, his reference system was not the apparently brightest group
member, but the unweighted average velocity of members. \citet{Zaritsky:1992p516}
studied asymmetric distribution of satellite galaxy velocities with Monte Carlo
simulations and concluded that observational biases partially explain the
observed redshift asymmetry. Despite the number of explanations none of
these explanations are satisfactory. Even new physics has been
suggested for the solution \citep[see e.g.][]{1970Natur.225.1033A}.

However, there may be a simple explanation of redshift asymmetries which
does not require modifications of well accepted physics, now that the
$\Lambda$CDM model can be counted as generally accepted. This was first proposed
by \citet{Byrd:1985p342} who pointed out that redshift asymmetries should arise
in nearby groups of galaxies such as HG groups if a large fraction of the group
population is unbound to the group. They argued that the redshift asymmetry
explains the need for "missing matter", the dark matter that was at the time
supposed to exist at the level of the closing density of the universe in groups
of galaxies. If the group as a whole is not virialized, there is no need for
excessive amounts of binding matter. It has now become possible to test this
assumption quantitatively, and this is the main focus of this paper. Independent
evidence has recently appeared of unbound outlying populations of galaxies around
the Local Group and a few other nearby groups, as one would expect in the
$\Lambda$CDM cosmology \citep{Teerikorpi:2008p589}. Thus in principle the
redshift asymmetry explanation of \citet{Valtonen:1986p340} should work; whether it
works quantitatively is a question to be answered in this paper.

In this paper we study redshift asymmetries theoretically in small groups of
galaxies by taking advantage of the largest cosmological $N-$body simulation
conducted so far: the Millennium Run \citep{Springel:2005p595}. We create two
'mock' catalogues of groups of galaxies from the semi-analytical galaxy catalogue
\citep{DeLucia:2007p414} of the Millennium Simulation by mimicking observational
methods. For the creation of mock catalogues we apply the Friends-of-Friends
percolation algorithm developed by HG82. The two mock catalogues differ in the
values of free parameters of the FoF grouping algorithm.

We show that the excess of positive redshifts is mainly due to wrong
identification of the dominant galaxy in a group and, at the same time, the group
being gravitationally unbound. We also show that groups that show a large excess
of positive redshifts are more often gravitationally unbound than groups that do
not show any significant excess. These errors in the identification of the
dominant galaxy result from our current inability to measure the relative
distances inside groups of galaxies, except for a few of the nearest ones
\citep{Karachentsev:1997p584,Jerjen:2001p583,Rekola:2005p586,Rekola:361p330R,Teerikorpi:2008p589}.
Due to peculiar motions of group members we cannot transform the apparent
magnitudes of galaxies in group catalogues into absolute magnitudes precisely. If
it were possible and the dominant group members were correctly identified, would
it lead to a small blueshift excess, which is due to the magnitude limited
observations that cause some of the background galaxies to be invisible and thus
excluded from groups.

This paper is organized as follows. In Section \ref{sample}, we discuss our
sample of galaxy groups, the Millennium Simulation data and the grouping
algorithm adopted. We present our findings and results in Section \ref{analysis}.
Finally, we summarize our results in Section \ref{summary}. Throughout this paper
we adopt a parametrized Hubble constant: $H_{0} = 100h$ km s$^{-1}$ Mpc$^{-1}$.
Unless explicitly noted, we adopt $h = 1.0$ for convenient comparison with older literature.

\section[]{The sample of galaxy groups}\label{sample}

The two mock catalogues of groups of galaxies from Millennium Simulation data were
constructed so that they would be comparable to the real observational group
catalogues HG82 and UZC-SSRS2 \citep{Ramella:2002p514} as much as possible. Both
of these catalogues are produced with the FoF algorithm. However, slightly
different values for the free parameters of the algorithm have been used. We show
in Section \ref{comparison} that our groups from the Millennium Simulation are
comparable to observed groups in a statistical sense. We also discuss briefly how
our results compare to works of other authors.

\subsection[]{Millennium Simulation data}\label{}

The Millennium Simulation \citep[MS;][]{Springel:2005p595} is a cosmological
$N-$body simulation of the $\Lambda$CDM model performed by the Virgo Consortium. The
MS was carried out with a customized version of the GADGET2 code developed by
\citet{Springel:2001p591}. The MS follows the evolution of $2160^{3}$
particles from redshift $z = 127$ in a box of 500h$^{-1}$ Mpc on a side. The
cosmological parameters of the MS simulation are: $\Omega_{m} = \Omega_{dm} +
\Omega_{b} = 0.25$, $\Omega_{b} = 0.045$, $h = 0.73$, $\Omega_{\Lambda} = 0.75$,
$n = 1$, and $\sigma_{8} = 0.9$ \citep[for a detailed description of the MS
see][]{Springel:2005p595}.

The galaxy formation modeling of the MS data is based on merger trees built from
$64$ individual snapshots. Properties of galaxies in MS data are obtained by
using semi-analytic galaxy formation models, where the star formation and its
regulation by feedback processes is parametrized in terms of analytical physical
models. A detailed description of the creation of the MS galaxy catalogue can be found in \citet{DeLucia:2007p414}, see also \citet{Croton:2006p249}.

The MS galaxy database does not directly give a morphology for galaxies. We have
used a method which takes an advantage of bulge-to-disk ratios to assign a
morphology to every galaxy. \citet{Simien:1986p431} found a correlation between
the B-band bulge-to-disc ratio, and the Hubble type $T$ of galaxies. The mean
relation may be written:
\begin{equation}\label{eq:T}
<\Delta m(T)> \quad = \quad 0.324x(T) - 0.054x(T)^{2} + 0.0047x(T)^{3} ,
\end{equation} 
where $\Delta m(T)$ is the difference between the bulge magnitude and the total
magnitude and $x(T) = T+5$. We classify galaxies with $T < -2.5$ as ellipticals,
those with $-2.5 < T < 0.92$ as S0s, and those with $T > 0.92$ as spirals and
irregulars. Galaxies without any bulge are classified as type $T = 9$. These
classification criteria are the same as proposed and adopted by
\citet{Springel:2001p363}.

\subsection[]{Group catalogues}\label{}

Our mock catalogues of groups of galaxies are generated with the FoF percolation
algorithm developed by HG82. Even though new algorithms have been developed, the
FoF still remains the most applied one. The FoF algorithm uses only two criteria
for finding group members: position and redshift. It essentially finds density
enhancements in position and in redshift space above a set threshold factor. This
threshold depends on a chosen value of the free parameter $D_{0}$, the apparent
magnitude limit of the search and the \citet{Schechter:1976p592} luminosity function. Density
enhancement relative to the mean number density can be calculated from the equation:
\begin{equation}
\frac{\delta \rho}{\rho} = \frac{3}{4\pi D_{0}^{3}}\left
(\int_{-\infty}^{M_{lim}} \Phi(M)dM \right)^{-1} - 1 ,
\end{equation}
where $D_{0}$ is the projected separation in Mpc chosen at some fiducial
redshift, $M_{lim} = m_{lim} - 30$ and $\Phi(M)$ is the \citet{Schechter:1976p592} luminosity
function. For a more detailed description of the FoF algorithm, see Paper I and
the references therein, especially HG82.

We produce two mock catalogues with different choices of free parameters. Both
group catalogues are generated from five independent volumes of the Millennium
Simulation galaxy catalogue. Each of the cubes used have a side length of
$250h^{-1}$ Mpc, and they do not overlap. The observation point inside
each volume was chosen to be in the centre of the particular cube. No additional
criteria were applied for the selection of observation points. Reasonable
statistical agreement, as shown in the next section between mock catalogues and
observed group catalogues shows that our method of choosing the origin without any
further criteria is strict enough in a statistical study of galaxy groups. Both
of our catalogues from simulations, Mock1 and Mock2, contain groups of galaxies
whose lower limit on the number of members, $n$, is 3. All groups containing $n >
2$ members are considered when group properties are studied in Section
\ref{comparison}. In Section \ref{redshift} we limit the number of group
members to $2 < n < 11$, comparable to S84.

Our first mock catalogue, named Mock1, is generated with the same values
( $m_{lim} = 13.2$ [in the Zwicky or de Vaucouleurs B(0) magnitude system], $D_{0} =
0.60~h^{-1}$ Mpc and $V_{0} = 400$ km s$^{-1}$) of the free parameters as the
original HG82 catalogue. This choice guarantees that we can compare results found
in S84 directly to our simulated catalogue and we can be sure that different
choices of parameter values do not effect the results. Even though the Mock1
catalogue uses the same values of parameters as HG82 it contains over 10 times
more groups than the original HG82 catalogue. Thus it provides significantly
better statistics.

Our Mock2 catalogue has an apparent B-band magnitude limit of $m_{lim} = 14.0$,
while $V_{0} = 200$ km s$^{-1}$ and $D_{0} = 0.37~h^{-1}$ Mpc were adopted for
the free parameters of the FoF algorithm, corresponding to the space density
enhancement of $\sim 68$. Despite the use of more strict parameters, Mock2
contains almost 10 times more groups than Mock1 due to the fainter apparent
B-band magnitude limit adopted. Because of the greater number of groups, Mock2 is
used for comparison and for better statistics. It should also contain groups
which are more often gravitationally bound due to the higher density enhancement
of groups.

Table \ref{tb:FOFparam} shows the values of free parameters used in creating our
mock catalogues. It also shows parameter values of various group catalogues based
on modern redshift surveys. All catalogues shown in Table \ref{tb:FOFparam} have
been generated with the FoF algorithm. However, catalogues based on the SDSS and
the 2dF data have taken advantage of modified versions of the original FoF.
The most noticeable modifications include the use of dark matter mock catalogues and group
re-centering.

\begin{table*}
\caption{Values of free parameters used for creating mock and observed group
catalogues.}
\label{tb:FOFparam}
\begin{tabular}{lllrrllll}
  \hline
  \hline
  Authors & Sample & $m_{lim}$ & $D_{0}$ & $V_{0}$ & $\frac{\delta \rho}{\rho}$
  & $\alpha$ & $M^{*}$ & $\phi^{*}$\\
  \hline 
  This paper & Mock1 & 13.2 & 0.64 & 400 & 19.8 & $-1.02$ & $-19.06$ & 0.0277\\
  This paper & Mock2 & 14.0 & 0.37 & 200 & 68.0 & $-1.15$ & $-19.84$ & 0.0172\\
  Huchra \& Geller (1982) & HG82 & 13.2 & 0.64 & 400 & 20 & $-1.02$ & $-19.06$ &
  0.0277\\
  Ramella et al. (2002) & UZC-SSRS2 & 15.5 & 0.25 & 350 & 80 & $-1.1~(-1.2)$ &
  $-19.1~(-19.73)$ & 0.04 (0.013)\\
  Eke et al. (2004) & 2dFGRS & 19.45 & 0.13 & 143 & $-$ & $-$ & $-$ & $-$\\
  \cite{Goto:2005p517} & SDSS-DR2 & 17.77 & 1.5 & 1000 & $-$ & $-$ & $-$ & $-$\\
  Merch\'an \& Zandivarez (2005) & SDSS-DR3 & 17.77 & $-$ & 200 & 80 & $-1.05$
  & $-20.44$ & $-$\\
  Berlind et al. (2006) & SDSS-DR3 & 17.5 & 0.14 & 75 & $-$ & $-$ & $-$ & $-$\\ 
  Tago et al. (2006) & 2dFGRS & 19.45 & 0.25 & 200 & $-$ & $-1.21$ & $-19.66$ &
  $-$\\
  Tago et al. (2008) & SDSS-DR5 & 17.77 & 0.25 & 250 & $-$ & $-$ & $-$ & $-$\\
  \hline
\end{tabular}
\medskip

Columns: The group catalogue used, $m_{lim}$ is the apparent magnitude
limit of the search (B-band for mock catalogues, r-band for the SDSS and
b$_{\textrm{j}}$ for the 2dF, note however that different redshift limits have
been applied in individual papers), $D_{0}$ is the projected separation in
$h^{-1}$ Mpc chosen at some fiducial redshift, $V_{0}$ is the velocity difference
in km s$^{-1}$, $\frac{\delta \rho}{\rho}$ is the density enhancement and
$\alpha$, $M^{*}$, and $\phi^{*}$ parametrize the \citet{Schechter:1976p592} function. The
Schechter parameters for the Mock2 sample are derived from the Millennium Galaxy
Catalogue by \cite{Driver:2007p587}. The Schechter parameters for Ramella et al. (2002)
refer to the CfA (SSRS2) groups.
\end{table*}

There are large differences between the Mock1 and Mock2 catalogues. The most obvious
difference is in the total number of groups. Mock1 contains
1601 groups in total while Mock2 contains 13786 groups. Note that five
different 'observation' points inside the MS are used, and none of these
observation points overlap each other. The difference in the number of groups is
due to the difference in the adopted apparent B-band magnitude limit. If we
compare the fraction of gravitationally bound groups between the two catalogues, (i.e. $T_{kin}/U < 1$, where
$T_{kin}$ is the kinetic energy and $U$ is the absolute value of the potential
energy of the group; for a detailed description see Paper I)
the differences are not great. The fraction of gravitationally bound
groups is surprisingly low in both mock catalogues. In the Mock1 catalogue the
fraction of bound groups is $52.64 \pm 9.21 \%$ while in Mock2 it is only
$50.19 \pm 1.15 \%$. The error limits are standard deviation errors
between the five observation points.

It is an indication of the reliability of the group-finding algorithm that the
same relative number of bound groups are found in spite of the fact that the Mock2
catalogue has about three times higher density enhancement than Mock1. In
Paper I we found that the fraction of gravitationally bound groups of dark matter
haloes 
is $\sim 30 \%$ when a $\Lambda$CDM model has been adopted. The fraction of
gravitationally unbound groups is significant in all these mock group catalogues,
which suggests that group catalogues based on the FoF algorithm contain a
significant fraction of groups that are not gravitationally bound systems.

\section[]{Analysis of groups}\label{analysis}

\subsection[]{Comparison with observations}\label{comparison}

In this subsection we briefly show that our mock catalogues are comparable to
real observational group catalogues. We compare our simulated mock catalogues to
UZC-SSRS2 and to HG82 catalogues. We only show comparisons in velocity
dispersion and in the 'observable' mass of groups. Even though our mock catalogues are
comparable to observations, we find some differences. We also find differences
between the two mock catalogues. Before discussing these differences we briefly
review the comparison catalogues and parameters.

The HG82 group catalogue was derived from a whole sky catalogue of 1312 galaxies
brighter than $m_{B} = 13.2$ (in Zwicky or de Vaucouleurs B(0) magnitude system)
with complete redshift information. The velocity of each galaxy has been
corrected for a dipole Virgo-centric flow. The catalogue of groups was obtained
with $D_{0} = 0.60~h^{-1}$ Mpc (corresponding to a density enhancement of
$\sim 20$) and with $V_{0} = 400$ km s$^{-1}$. Only groups containing more than
two members have been included in the final catalogue.

The UZC-SSRS2 group catalogue was derived from a magnitude-limited redshift
sample of galaxies. A compilation of 6846 galaxies with the apparent magnitude limit of
$m_{lim} \leq 15.5$ was used for the creation of the UZC-SSRS2 catalogue, which
contains, in total, 1168 groups. The group catalogue covers 4.69 sr, and the
parameter values of $V_{0} = 350$ km s$^{-1}$ and $D_{0} = 0.25~h^{-1}$ Mpc have
been adopted for the creation of the UZC-SSRS2 catalogue. These values correspond
to a density contrast threshold $\sim 80$. Only groups containing more than two
members have been included in the final catalogue.

Our mock catalogues were discussed in the previous Section. However, we would
like to point out that our Mock1 (Mock2) catalogue is comparable to HG82
(UZC-SSRS2) in the choice of parameters. The Mock2 catalogue does not adopt exactly
the same parameters as the UZC-SSRS2, even though the density enhancement is
comparable. The reason for not adopting exactly the same values is that volumes
inside the MS are not large enough. Adopting $m_{lim} = 15.5$ would have
introduced errors in groups and their properties due to edge effects and missing
group members. We wish to avoid this, as our purpose is to study distribution of
group members and possible redshift asymmetries.

We have shown in Paper I that cosmological $N-$body simulations can produce
groups of galaxies which are comparable to observations. However, in Paper I, we
were greatly limited by the volume of our simulation boxes, causing comparisons
to be less conclusive. Moreover, In Paper I we compared properties of dark
matter haloes to real observed galaxies. As Millennium Simulation offers a larger
volume and the properties of the galaxy data are derived with semi-analytical
models, the comparison between mock catalogues and observational catalogues
(UZC-SSRS2 and HG82) is now more robust. We use the same definitions and
equations for velocity dispersion and 'observable' mass of a group as in Paper I.

The velocity dispersion $\sigma_{v}$ of a group is defined as: 
\begin{equation} 
	\sigma_{v} = \sqrt{ \frac{1}{N_{G} - 1} \sum_{i = 1}^{N_{G}}(v_{i} -
	<v_{R}>)^{2}} ~ , 
\end{equation} 
where $N_{G}$ is the number of galaxies in a group, $v_{i}$ is the radial
velocity of the $i$th galaxy and $<v_{R}>$ is the mean group radial velocity. In
observations, the group masses can be estimated by various methods. In the HG82
and in the UZC-SSRS2 catalogues the total mass of a group is estimated with a
simple relation:
\begin{equation}\label{obsmass}
	M_{obs} = 6.96\times10^{8}\sigma_{v}^{2}R_{H}M_{\odot} ~ ,
\end{equation}
where $\sigma_{v}$ is the velocity dispersion, and $R_{H}$ is the mean harmonic 
radius:
\begin{equation} 
	R_{H} = \frac{\pi <v_{R}>}{H_{0}} \sin \left\lbrace \frac{1}{2} 
	\left[ \frac{N_{G}(N_{G}-1)}{2} \left( \sum_{i=1}^{N_{G}} \sum_{j>i}^{N_{G}} 
	\theta_{ij} \right)^{-1} \right] \right\rbrace  ~ , 
\end{equation} 
where $\theta_{ij}$ is the angular separation of the $i$th and $j$th group members.

\begin{figure}
\includegraphics[width=84mm]{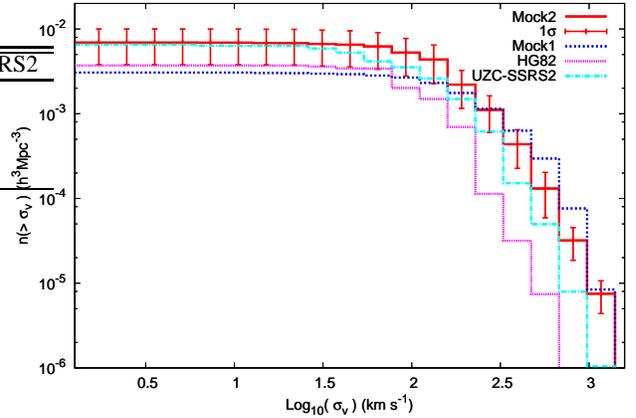}
\caption{The cumulative number density of velocity dispersion $\sigma_{v}$ for
galaxy groups. Simulation data are  averaged over the ensemble of 5 observation
points. The error bars are 1$\sigma$ errors and are only shown for Mock2 for
clarity. The error bars for other data have similar or slightly greater size due
to poorer statistics.}
	\label{F:velocitydisp}
\end{figure}

\begin{figure}
\includegraphics[width=84mm]{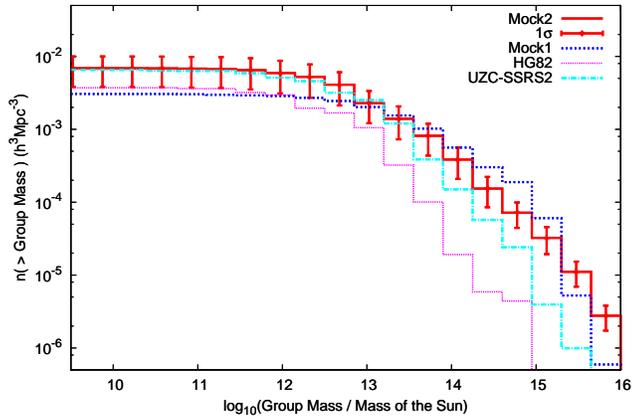}
\caption{Group abundance by 'observable' mass of the groups. Simulation data are
averaged over the ensemble of 5 observation points.  The error bars are 1$\sigma$
errors and are only shown for Mock2 for clarity. The error bars for other data
have similar or slightly greater size due to poorer statistics.}
	\label{F:obsmass}
\end{figure}

To compare abundances of groups in magnitude-limited samples we weight each group
according to its distance \citep{Moore:1993p360,Diaferio:1999p398}. This
weighting is necessary as there is no 'total' volume of a galaxy sample in
magnitude-limited group catalogues. After weighting each group individually we
can scale abundances of groups in Figs. \ref{F:velocitydisp} and \ref{F:obsmass}
to the comoving volume of the sample. We include all galaxies with $cz > 500$ km
s$^{-1}$. This lower cut-off avoids including faint objects that are close to the
observation point as these groups could contain galaxies fainter than in real
magnitude-limited surveys. Therefore we consider only groups with mean radial
velocity, $<cz>$, greater than $500$ km s$^{-1}$ in Mock1, Mock2, HG82 and
UZC-SSRS2 catalogues.

Figs. \ref{F:velocitydisp} and \ref{F:obsmass} show that cosmological $N-$body
simulations can produce groups of galaxies whose statistical properties are
similar to observed ones. The agreement is, in general, within 2$\sigma$ error
bars. However, there are clear differences visible between all catalogues in both
Figures. These differences are discussed below in greater detail. We use a
statistical Kolmogorov-Smirnov (K$-$S) test to prove or disprove the null
hypothesis, $H_{null}$, that the two distributions are alike and are drawn from
the same population distribution function. Results of the K$-$S tests are
presented as significance levels (value of the $Q$ function) for the null
hypothesis and are listen in Table \ref{tb:KS}.

\begin{table}
\caption{The significance levels of the K$-$S tests.}
\label{tb:KS}
\begin{tabular}{lccccr}
  \hline
  \hline
  Sample & Property & HG82 & UZC-SSRS2 & HG82 vs. UZC-SSRS2\\
  \hline
  Mock1 & $\sigma_{v}$	&$0.013$			& $2.7\times 10^{-4}$ 	& $0.08$\\
  Mock1 & $M_{G}$ 		&$4.2\times 10^{-4}$& $7.1\times 10^{-7}$	& $0.05$\\
  Mock2 & $\sigma_{v}$	&$4.8\times 10^{-4}$& $0.015$ 				& $0.08$\\
  Mock2 & $M_{G}$ 		&$0.042$			& $0.462$				& $0.05$\\
  \hline
\end{tabular}

\medskip
Note: Significance levels of the K$-$S test for the null hypothesis that
observations and the simulations are alike and are drawn from the same parent
population (HG82 and UZC-SSRS2 columns). Significance level of the K$-$S test for
the null hypothesis that the HG82 and the UZC-SSRS2 group catalogue are alike and
are drawn from the same parent population (HG82 vs. UZC-SSRS2 column).
\end{table}

Both Figs. show that the total density of the Mock1 and HG82 catalogue is lower than
of Mock2 and UZC-SSRS2. This result is due to the lower apparent magnitude limit
of these catalogues, even when weighting is applied. Mock1 and the HG82 are
missing more faint galaxies than Mock2 and UZC-SSRS2 due to the lower magnitude
limit. Because of this, these catalogues are missing groups that are built only
from relatively faint galaxies i.e. these catalogues miss small groups with small
velocity dispersions in comparison to Mock2 and UZC-SSRS2.

Fig. \ref{F:velocitydisp} shows that groups of galaxies in our mock catalogues
are similar to observed ones in a statistical sense when velocity dispersions are
studied. However, both mock catalogues show an excess of high velocity dispersion
groups in comparison to observations. Despite these differences the K$-$S test is
approved (at level of $0.01$) for Mock1 when the comparison is to HG82 and for
Mock2 when the comparison is to the UZC-SSRS2 catalogue. It is also noteworthy
that the Mock1 catalogue shows a small excess of groups around $\sim 500$ km s$^{-1}$
in comparison to Mock2. This difference is probably due to the larger value of
$V_{0}$ in Mock1 that allows a greater difference between group members in
redshift space.

Qualitatively better agreement in velocity dispersion is observed when mock
catalogues are compared to the UZC-SSRS2 catalogue. There are also some
differences between the HG82 and UZC-SSRS2 catalogue, especially when the abundance
of high velocity dispersion groups are considered. The discrepancy between the mock
catalogues and the HG82 group catalogue is relatively large when large velocity
dispersions are considered. This difference is due to the low number of groups
(92) in HG82. Also the volume of the HG82 catalogue is relatively small. Thus,
the HG82 catalogue lacks high velocity dispersion groups and clusters as the
Virgo cluster is the only big cluster within the visible volume of the catalogue.

The quartile values of velocity dispersion of the Mock1 (Mock2) catalogue
groups are 131.0/213.4/347.4 km s$^{-1}$ (80.5/131.0/213.4 km s$^{-1}$). The
values of the Mock2 catalogue groups are closer to the observed ones than the values
of mock catalogue groups of Paper I. The large values of Mock1 groups can be
explained with larger number of interlopers due to the lower density enhancement
and the relatively large value of $V_{0}$. The above differences show that the
selection of $V_{0}$ is important, especially for velocity dispersions of
groups.

From Fig. \ref{F:obsmass} it is clear that our mock catalogues show an excess in
the abundance of heavier groups. Despite the differences the K$-$S test supports
the null hypothesis for the Mock2 catalogue (for numerical details see Table
\ref{tb:KS}). Because of the strong connection between group mass and the
velocity dispersion (see Eq. \ref{obsmass}) the Mock1 catalogue shows (qualitatively)
similar behavior in both Figs. \ref{F:velocitydisp} and \ref{F:obsmass}. There is
also a significant difference between HG82 and UZC-SSRS2 catalogue when the
abundance of massive groups is studied. However, the most striking difference is
between mock catalogues and HG82 when massive groups and clusters are considered.
These differences can be explained with the small volume and the low number of
large groups in HG82. In total, HG82 has only two groups with more than 30 members
while the UZC-SSRS2 catalogue has 14 groups. The Mock1 (Mock2) catalogue has 45 (276)
groups that have more than 30 members. It is obvious that the most massive groups
are the ones that have highest velocity dispersion and that are the most expanded
ones, meaning simply the ones having most members.

The quartile values of 'observable' mass of the Mock1 (Mock2) catalogue groups
are $0.7/3.5/7.9 \times 10^{13}$M$_{\odot}$ ($1.4/3.1/15.8 \times
10^{12}$M$_{\odot}$). The values of Mock2 are close to the observed
values (see Table 4 in Paper I for numerical details) even if there is an excess in
the abundance of massive groups in comparison to observations. The good agreement
in quartile values suggests that simulated groups are similar in a statistical
sense, as the greatest differences in Fig. \ref{F:obsmass} are observed at very
low densities.

Despite the differences discussed above, our mock catalogues are comparable to
observational catalogues in a statistical sense (see Table \ref{tb:KS}). This
conclusion is supported by the K$-$S tests, by the good agreement on 
quartile values and by the fact that the properties of simulated groups are, in
general, within $2\sigma$ error limits. We do not compare other properties of
groups in this paper, as the focus of the paper is to study and explain the
observed redshift asymmetries. However, before explaining these asymmetries
we briefly compare our results to the findings of other authors.

\citet{Eke:2004p509, 2006MNRAS.370.1147E} have previously applied the same
method to galaxy groups identified in the 2dFGR Survey. These groups are
typically much further away than our groups, as the median redshift of their groups is $0.11$.
Despite this difference we make an attempt to compare their results to ours. \citet{Eke:2004p509} found the median velocity dispersion of
$227$ km s$^{-1}$ for groups with at least three members. This is rather close to
our Mock1 catalogue value ($\sim 213$ km s$^{-1}$) implying that our mock groups
are similar to their groups in a statistical sense. The median velocity dispersion
of the Mock2 catalogue is only $\sim 131$ km s$^{-1}$, and differs significantly from
the value of \citet{Eke:2004p509}. This result is somewhat expected, as our mock
groups are all found at very close distances. The median value of dynamical mass in
our Mock1 catalogue is close to the value of \citet{Eke:2004p509}; even though their
definition of dynamical mass is different to ours.

\citet{Eke:2004p509} found that as much as $\sim 40$ per cent of groups can
contain interlopers. This fraction is close to the fraction of groups we find to
be gravitationally unbound. \citet{2006MNRAS.370.1147E} found that dynamical
group masses give higher abundances to the mass function. It is possible that the
reason for this discrepancy lies in the large fraction of unbound groups. If
the (dynamical) mass of a group is calculated by adopting the virial theorem, one has
to assume that the group is a bound structure. However, if one applies the virial
theorem to a group that is gravitationally unbound, the dynamical mass of the
group can be overestimated significantly. Even if the virial theorem is not used
while calculating the dynamical mass, one can easily overestimate the mass, as
unbound groups can have significantly higher velocity dispersion and size,
leading to the higher abundance noticed by \citet{2006MNRAS.370.1147E}.

As groups observed by \citet{Eke:2004p509, 2006MNRAS.370.1147E} are typically much further
away, these groups are not expected to show significant redshift asymmetry. In
our groups the distance ratio of the far side of the group relative to the front
side of the group can be rather large, and this causes the interesting effects
that are discussed next.

\subsection[]{Redshift asymmetries in groups}\label{redshift}

In this subsection we use our mock catalogues to study and explain the observed
redshift asymmetries. However, before that, we discuss the suitability of our mock
catalogues and quantify the methods of calculating the redshift asymmetries.


The Mock1 catalogue has been generated with the same parameters as the original HG82
catalogue, from which S84 found positive redshift excess. Because of this it is
most suitable for this study and for the comparison of redshift excesses found in S84.
Mock1 is also favoured because a redshift excess is most noticeable for
nearby groups for which there is a larger distance ratio between the background
and the foreground galaxies. Mock2 is used to provide better statistics
and for comparison. Note, however, that Mock2 probes groups deeper in
redshift space than Mock1, and therefore the total effect of nearby
groups is not as strong. It is also noteworthy that in this subsection we study
small groups, therefore we limit the number of group members to less than 11
(i.e. $2 < n \leq 10$).

Both mock catalogues have been generated from the Millennium Simulation.
Even though the Millennium run provides reliable statistics its mass resolution is only
$\sim 10^{9}$M$_{\odot}$. This complicates the matter of identifying low mass
groups with at least three members inside. The low mass resolution might have a
large effect on studies of the abundance or space density of
individual objects. An even greater effect would be noticed if one was interested in
subhalo properties or abundances. However, as our purpose is to study the
relative location and distribution of galaxies inside groups, we do not consider
the low mass resolution as a significant problem. Further, Section
\ref{comparison} showed that the abundance of low mass groups in the Millennium
Simulation is comparable to observable catalogues. This ensures that the mass
resolution is good enough for a statistical study like ours, especially as in
a statistical sense missing dwarf galaxies could reside anywhere inside the dark
matter halo.

We calculate redshift asymmetries both on a group and on a galaxy level. On a
group level we use groups from five different observation points and calculate
the sum of groups with redshift/blueshift excess in each point. The redshift or
the blueshift excess percentage shown is the mean value of the excesses from the
five observation points when ties have been removed. The error limits are standard
deviation errors between five observation points. When the redshift asymmetries
are studied on a member galaxy level, we quantify the redshift excess as in
\citet{Byrd:1985p342}:
\begin{equation}\label{eq:Z}
Z = \frac{N_{R} - N_{B}}{N_{R} + N_{B}} ~ ,
\end{equation}
where $N_{B}$ is the number of galaxies having a redshift lower than the
apparently brightest group member and $N_{R}$ is the number of galaxies having
a redshift higher than the apparently brightest group member.

Table \ref{tb:comS1} shows a general comparison of the mock catalogues and
samples from S84 when the asymmetries have been calculated on a group level. Mock
catalogues show a (weak) positive redshift excess when spiral-dominated (i.e. the
central galaxy is a spiral galaxy) groups are studied. However, the excess is not
significant for elliptical (E or S0) dominated groups. These groups actually show
a small blueshift excess, similar to S84. By conventional theory, this suggests
that these groups could be mostly gravitationally bound. However, we find that
$\sim 35.0 \pm 10.0$ ($27.4 \pm 1.3$) per cent of E/S0-dominated groups in Mock1
(Mock2) are gravitationally unbound. This suggests that the absence of
redshift asymmetry does not alone guarantee that these groups are
gravitationally bound systems.


\begin{table*}
\caption{A general comparison of the mock catalogues and samples from S84.}
\label{tb:comS1}
\begin{tabular}{lccccccc}
  \hline
  \hline
  Sample & $n_{G}$ & $n_{+ \Delta z}$ & $n_{- \Delta z}$ & $n_{\pm \Delta z}$ &
  $+ \Delta z$ (\%) & $- \Delta z$ (\%) & $\pm \Delta z$ (\%)
  \\
  \hline 
  S84, all groups ($n \leq 10$) & 85 & 42 & 32 & 11 & 57.0 & 43.0 & 12.9\\ 
  S84, E/S0 dominant groups removed & 60 & 33 & 18 & 9 & 65.0 & 35.0 & 15.0\\
  S84, E/S0 dominant groups & 21 & 9 & 10 & 2 & 47.0 & 53.0 & 9.5\\
  \hline
  Mock1, all groups ($n \leq 10$) & 1384 & 622 & 505 & 257 & 55.2 $\pm$ 5.1 &
  44.8 $\pm$ 5.1 & 18.6 $\pm$ 1.9\\
  Mock1, spiral (T$ \geq 0.92$) dominant groups & 1153 & 528 & 408 & 217 & 56.4 $\pm$
  5.6 & 43.6 $\pm$ 5.6 & 18.8 $\pm$ 2.4\\
  Mock1, E/S0 (T$ < 0.92$) dominant groups & 231 & 94 & 97 & 40 & 49.2 $\pm$ 6.9
  & 50.8 $\pm$ 6.9 & 17.3 $\pm$ 4.2\\
  \hline
  Mock2, all groups ($n \leq 10$) & 12100 & 5087 & 4658 & 2355 & 52.2 $\pm$ 1.3
  & 47.8 $\pm$ 1.3 & 19.5 $\pm$ 0.6\\
  Mock2, spiral (T$ \geq 0.92$) dominant groups & 9086 & 3854 & 3466 & 1748 & 52.5
  $\pm$ 1.8 & 47.3 $\pm$ 1.6 & 19.2 $\pm$ 0.8\\
  Mock2, E/S0 (T$ < 0.92$) dominant groups & 3013 & 1179 & 1227 & 607 & 49.0
  $\pm$ 1.8 & 51.0 $\pm$ 1.8 & 20.1 $\pm$ 2.0\\
  \hline
\end{tabular}
\medskip 

Note: Sample refers either to S84, Mock1 or Mock2 catalogues, $n_{G}$ is the
number of groups, $n_{+ \Delta z}$ is the number of groups with positive redshift
excess, $n_{- \Delta z}$ is the number of groups with negative redshift excess,
$n_{\pm \Delta z}$ is the number of groups with no excess, $+ \Delta z$ is the
percentage of groups with positive redshift excess when ties have been removed, $-
\Delta z$ is the percentage of groups with negative redshift excess when ties have
been removed and $\pm \Delta z$ is the percentage of groups having equal number
of negative and positive redshifts. Errors for mock catalogues are standard
deviation errors between five different observation points.
\end{table*}

Table \ref{tb:detailed} further divides the mock catalogues to subsamples on the
basis of whether the groups are gravitationally bound or not. The unbound groups
are important, since about one half of all groups in both of our mock catalogues
belong to this category. A further division in Table \ref{tb:detailed} is made on
the basis of whether the dominant galaxy is correctly identified, i.e. whether
the brightest, in apparent magnitudes group member is also the most massive
galaxy in the group. A large and statistically highly significant redshift excess
appears only in those subsamples where the groups are gravitationally unbound,
and in addition, their dominant galaxies have been misidentified. These comprise
approximately one quarter of all groups. The excess appears both among elliptical
and spiral dominated groups but is stronger among the spiral dominated groups.

Table \ref{tb:detailed} also shows that groups that are
gravitationally bound show larger error limits for redshift asymmetries than
groups which are unbound, suggesting that there are large differences in the
fraction of bound groups between our five observation points. This result shows
that projection effects can play a significant role when the FoF percolation
algorithm is applied, and that it can produce groups which are spurious because
of these projection effects. It also shows that the choice of free parameters can
affect the results, as the Mock2 catalogue does not show larger errors for bound
groups in comparison to unbound ones.


When redshift asymmetries are studied at a member galaxy level, the results stay
similar. If we consider only spiral-dominated groups which are unbound and whose
centre is wrongly identified, we find a redshift excess of $Z = 0.22$ and $Z =
0.10$ for the Mock1 and Mock2 catalogues, respectively. When only bound groups from
Mock1 (Mock2) are considered, the redshift excess is only $0.07$
($0.04$). The former values are similar to observed ones, as S84 found a redshift
excess of $Z = 0.21$ for spiral-dominated (i.e. the apparently brightest galaxy
is a spiral galaxy) groups. \citet{Byrd:1985p342} and \citet{Valtonen:1986p340}
derived values of the redshift excess from an analytical model that ranges from $Z
= 0.1$ to $Z = 0.5$ depending on the parameters adopted, while the most probable
value was $0.2$. The good agreement between observations, analytical models and
simulations (unbound and wrongly identified groups) suggests that most of the
membership of HG82's spiral dominated groups is unbound and that the centre has
been misidentified for these groups. Simulations further show that to obtain as high
a redshift excess as observed, we have to select groups that are gravitationally
unbound and whose centre has been misidentified.

If we concentrate on E/S0-dominant groups at a member galaxy level, we see
similar results as in Tables \ref{tb:comS1} and \ref{tb:detailed}. Now Mock1
(Mock2) shows a blueshift excess of $Z = -0.05$ ($Z = -0.01$) for
gravitationally unbound groups, whose central galaxies are correctly identified.
This is a slightly weaker blueshift excess than S84 found ($-0.13$), but
comparable. These values are in agreement with the analytical calculations of
\citet{Byrd:1985p342} and \citet{Valtonen:1986p340} that showed the blueshift excess of correctly
identified, but unbound E-dominated groups ranging from $Z = -0.13$ to $Z = 0.03$
depending on the values of free parameters. Here the agreement between observations,
analytical models and simulations is reasonable. This points to the possibility
that E/S0 galaxies do mark the position of the group centre correctly.

Table \ref{tb:detailed} shows that E/S0 galaxies mark the position of the centre
correctly more often than spiral galaxies. If we calculate the percentage of
groups whose dominant member has been correctly identified it is clear that $\sim
73.9$ per cent of E/S0-dominated groups are correctly identified while as much
as $\sim 60.0$ per cent of spiral-dominant groups are incorrectly identified. The
incorrect identification of the central galaxy is important as it may lead to a
redshift excess, since it is more likely that the apparently brightest galaxy is
in the front part of the cluster than in the back part of it. The galaxies in the
front part of the group appear brighter, and in the back part of the group
fainter than what their absolute magnitudes would lead us to expect, i.e.
compared with the situation when they all are at the same distance from us. This
greater apparent brightness of the group members in the front makes it also more
likely that the apparently brightest member is picked from the front volume of
the group. Statistically this applies to groups of all sizes. In a
gravitationally bound group this would not in itself cause a redshift excess, but
if the whole group or a substantial part of it is in Hubble flow, then the
galaxies in front are typically blueshifted while the galaxies in the back are
redshifted relative to the apparently brightest galaxy.

\begin{table*}
\caption{A detailed view of the redshift asymmetries in mock catalogues.}
\label{tb:detailed}
\begin{tabular}{lccccccr}
  \hline
  \hline
  Sample & $n_{G}$ & $n_{+ \Delta z}$ & $n_{- \Delta z}$ & $n_{\pm \Delta z}$ &
  $+ \Delta z$ (\%) & $- \Delta z$ (\%) & $\pm \Delta z$ (\%)
  \\
  \hline 
  Mock1, gravitationally bound groups\\
  all groups ($n \leq 10$) & 675 & 282 & 259 & 134 & 50.5 $\pm$ 10.1 & 49.5
  $\pm$ 10.1 & 19.9 $\pm$ 2.5\\
  spiral (T$ \geq 0.92$) dominant groups & 537 & 231 & 199 & 107 & 52.4 $\pm$
  10.6 & 47.6 $\pm$ 10.6 & 19.9 $\pm$ 2.7\\
  E/S0 (T$ < 0.92$) dominant groups & 138 & 51 & 60 & 27 & 46.0 $\pm$ 16.0 &
  54.1 $\pm$ $16.0$ & 19.6 $\pm$ 4.4\\
  \hline
  Mock1, gravitationally unbound groups\\
  all groups ($n \leq 10$) & 709 & 340 & 246 & 123 & 58.3 $\pm$ 2.9 & 41.7 $\pm$
  2.9 & 17.3 $\pm$ 2.2\\
  spiral (T$ \geq 0.92$) dominant groups & 616 & 297 & 209 & 110 & 59.0 $\pm$ 3.2 & 41.0
  $\pm$ 3.2 & 17.9 $\pm$ 2.8\\
  E/S0 (T$ < 0.92$) dominant groups & 93 & 43 & 37 & 13 & 54.1 $\pm$ 7.1 & 45.9
  $\pm$ 7.1 & 14.0 $\pm$ 8.5\\
  \hline
  Mock1, bound, wrongly identified\\
  all groups ($n \leq 10$) & 289 & 142 & 110 & 37 & 56.3 $\pm$ 12.9 & 43.7
  $\pm$ 12.9 & 12.8 $\pm$ 4.2\\
  spiral (T$ \geq 0.92$) dominant groups & 257 & 129 & 95 & 33 & 57.6 $\pm$ 12.8 & 42.4
  $\pm$ 12.8 & 12.8 $\pm$ 3.7\\
  E/S0 (T$ < 0.92$) dominant groups & 32 & 13 & 15 & 4 & 46.4 $\pm$ 18.2 & 53.6
  $\pm$ 18.2 & 12.5 $\pm$ 12.0\\
  \hline
  Mock1, unbound, wrongly identified\\
  all groups ($n \leq 10$) & 338 & 193 & 96 & 49 & 66.8 $\pm$ 3.8 & 33.2 $\pm$
  3.8 & 14.5 $\pm$ 4.6\\
  spiral (T$ \geq 0.92$) dominant groups & 297 & 171 & 83 & 43 & 67.6 $\pm$ 3.5 & 32.4
  $\pm$ 3.5 & 14.5 $\pm$ 4.8\\
  E/S0 (T$ < 0.92$) dominant groups & 41 & 22 & 13 & 6 & 62.9 $\pm$ 9.7 & 37.1
  $\pm$ 9.7 & 14.6 $\pm$ 9.9\\
  \hline
  Mock2, gravitationally bound groups\\
  all groups ($n \leq 10$) & 5715 & 2327 & 2214 & 1174 & 51.3 $\pm$ 1.7 & 48.7
  $\pm$ 1.7 & 20.5 $\pm$ 0.7\\
  spiral (T$ \geq 0.92$) dominant groups & 3649 & 1515 & 1386 & 748 & 52.3 $\pm$ 2.9
  & 47.7 $\pm$ 2.9 & 20.5 $\pm$ 1.2\\
  E/S0 (T$ < 0.92$) dominant groups & 2066 & 801 & 839 & 426 & 48.8 $\pm$ 2.2 &
  51.2 $\pm$ 2.2 & 20.6 $\pm$ 2.5\\
  \hline
  Mock2, gravitationally unbound groups\\
  all groups ($n \leq 10$) & 6385 & 2760 & 2444 & 1181 & 52.9 $\pm$ 1.6 & 47.1
  $\pm$ 1.6 & 18.5 $\pm$ 0.9\\
  spiral (T$ \geq 0.92$) dominant groups & 5437 & 2357 & 2080 & 1000 & 53.0 $\pm$
  1.7 & 47.0 $\pm$ 1.7 & 18.4 $\pm$ 1.1\\
  E/S0 (T$ < 0.92$) dominant groups & 947 & 402 & 364 & 181 & 52.4 $\pm$ 2.8 &
  47.6 $\pm$ 2.8 & 19.1 $\pm$ 1.7\\
  \hline
  Mock2, bound, wrongly identified\\
  all groups ($n \leq 10$) & 2618 & 1193 & 1028 & 397 & 53.7 $\pm$ 2.2 & 46.3
  $\pm$ 2.2 & 15.2 $\pm$ 0.7\\
  spiral (T$ \geq 0.92$) dominant groups & 2151 & 980 & 840 & 331 & 53.8 $\pm$ 2.7 &
  46.2 $\pm$ 2.7 & 15.4 $\pm$ 1.0\\
  E/S0 (T$ < 0.92$) dominant groups & 467 & 213 & 188 & 66 & 53.1 $\pm$ 1.7 &
  46.9 $\pm$ 1.7 & 14.1 $\pm$ 3.4\\
  \hline
  Mock2, unbound, wrongly identified\\
  all groups ($n \leq 10$) & 3591 & 1661 & 1379 & 551 & 54.6 $\pm$ 1.9 & 45.4
  $\pm$ 1.9 & 15.3 $\pm$ 1.3\\
  spiral (T$ \geq 0.92$) dominant groups & 3291 & 1535 & 1260 & 496 & 54.8 $\pm$ 2.0 &
  45.2 $\pm$ 2.0 & 15.1 $\pm$ 1.4\\
  E/S0 (T$ < 0.92$) dominant groups & 300 & 126 & 119 & 55 & 51.4 $\pm$ 5.6 &
  48.6 $\pm$ 5.6 & 18.3 $\pm$ 4.1\\
  \hline
\end{tabular}
\medskip 

Note: Sample is the subsample of the mock catalogue, $n_{G}$ is the number of
groups, $n_{+ \Delta z}$ is the number of groups with positive redshift excess,
$n_{- \Delta z}$ is the number of groups with negative redshift excess, $n_{\pm
\Delta z}$ is the number of groups with no excess, $+ \Delta z$ is the percentage
of groups with positive redshift excess when ties have been removed, $- \Delta z$
is the percentage of groups with negative redshift excess when ties have been
removed and $\pm \Delta z$ is the percentage of groups having equal number of
negative and positive redshifts. Errors for mock catalogues are standard
deviation errors between five different observation points.
\end{table*}

To further test the cause of redshift asymmetries we use absolute B-band
magnitudes rather than apparent ones to identify the dominant group member and
calculate the redshift asymmetries in this case. The use of absolute magnitudes
leads to a small blueshift excess for all groups. The blueshift excess is highest
for unbound, spiral-dominated and misidentified groups, being $\sim 56 \pm 1.5$
per cent ($Z \sim -0.06$). For all gravitationally bound groups we find a small
blueshift excess of $\sim 52.5 \pm 2.5$ ($Z \sim -0.02$). This small and hardly
significant blueshift excess is expected and is due to the fact that
magnitude limited observations will miss faint members from the back part of the
group with higher probability than from the front part. Additionally, if the group is
expanding (unbound) it can further lead to a case where the group is missing
redshifted members. A small blueshift excess is also present if we consider only
groups that are correctly identified.

Although the redshift excess disappears when using absolute B-band magnitudes, it
does not lead to a significantly smaller number of misidentifications as the
fraction remains roughly the same as in the case of apparent magnitudes. The
percentage of misidentifications for all galaxies in the case of apparent magnitudes
for Mock2 is $\sim 52.6\pm1.7 \%$ while it is $\sim 51.8\pm1.8 \%$
in the case of absolute magnitudes. This result is due to the fact that both
B-band magnitudes are a poor indicator of the most massive
galaxy as they tend to favour spiral galaxies over elliptical ones.

\section[]{Summary and conclusions}\label{summary}

In this paper we have compared cosmological $N-$body simulations to observations.
We have studied fractions of unbound groups, redshift asymmetries and their
connection. We have found an explanation for the positive redshift excess found by
many authors from different observational group catalogues.

Our mock catalogues of groups of galaxies are in reasonable agreement (in
general, within 2$\sigma$) with observational catalogues when dynamical
properties of groups are studied. Both mock catalogues show a significant
fraction of gravitationally unbound groups, independent of the choices of free
parameter values in the Friends-of-Friends algorithm. Even though the density
enhancement of the Mock2 catalogue is more than three times higher than the density
enhancement of Mock1, we found that both mock catalogues contain roughly $50 \%$
gravitationally unbound groups. Even though the density enhancement does
not have a significant effect on the fraction of gravitationally bound groups,
the values of free parameters of the FoF algorithm have an effect on group
properties. Our results show that the value of $V_{0}$ has an effect on the
abundance of high velocity groups. This is an expected result as a higher $V_{0}$
value gives the percolation algorithm more room in redshift space leading to a
higher number of high velocity dispersion groups.

Mock catalogues produce similar redshift asymmetries as found in observations.
The Mock1 catalogue produces higher redshift asymmetries than Mock2 in all
cases, because the redshift asymmetries are most noticeable for nearby groups for
which there is a larger distance ratio between the background and the foreground
galaxies. The redshift excess is similar to observations for groups that are
gravitationally unbound and in which, at the same time, the apparently brightest
galaxy is not the most massive galaxy. The misidentification of the group centre
is important as it can lead to a redshift excess, since it is more likely that
the apparently brightest galaxy is in the front part of the cluster than in the
back part of it. The galaxies in the front part of the group appear brighter, and
in the back part of the group fainter than what their absolute magnitudes would
lead us to expect, i.e. compared with the situation when they all are at the same
distance from us.

The use of absolute B-band magnitudes does not lead to a redshift excess; however,
a small blueshift excess is present. This is due to the fact that magnitude
limited observations miss faint group members from the back rather than from the
front part of the group. Despite the lack of redshift excess, the fraction of
groups in which the dominant group member has been incorrectly identified remains
as we observe roughly the same number of misidentified groups ($\sim 52$ per
cent) as in the case of apparent magnitudes when all groups from Mock2 are
considered. Thus, absolute and apparent B-band magnitude is a poor indicator of
the dominant member in a group. Our results also show that the E/S0 galaxies tend
to mark the group centre correctly, as $\sim 75$ per cent of E/S0-dominated
groups have been correctly identified. These groups do not show significant
redshift excess in any case.

Gravitationally bound groups do not show any significant redshift excess. This is
in agreement with conventional theory, where it is expected that
distribution of redshift differentials should be evenly distributed. The
subsample of the Mock1 catalogue that includes only gravitationally bound groups has
an equal number of galaxies relative to the brightest member within statistical
errors.

We conclude that when the group centre is not correctly identified, it may cause
the major part of the observed redshift excess. If the group is also
gravitationally unbound, the level of the redshift excess becomes as high as in
S84. Thus the explanation of \citet{Byrd:1985p342} and \citet{Valtonen:1986p340}
for the origin of the redshift excess is verified. It further means that there is
no need to introduce any ``anomalous'' redshift mechanism to explain the redshift
excess of \citet{1970Natur.225.1033A}.
 

This paper shows that the Friends-of-Friends percolation algorithm picks out the
expanding parts of the groups, in addition to the gravitationally bound group
cores. Thus the group catalogues constructed in this way cannot be used as if the
groups are purely bound systems. For example, the use of the virial theorem to
estimate group masses easily leads to wrong answers. As about 50 per cent of
groups in our mock catalogues are bound, in principle one could apply the virial
theorem only to this subclass of groups, but then it is difficult to identify
this subclass in observations. The redshift excess in a sample of groups would
tell us readily that there must be many unbound groups in the sample. However,
the absence of redshift excess alone does not guarantee that these groups are
gravitationally bound systems.
	
To overcome the difficulty of finding gravitationally bound groups, one can, for
example, use stellar mass rather than luminosity for identifying the central
galaxy. Even then dark matter and the lack of knowledge of relative distances
inside observed groups complicates matters. If the grouping algorithm
concentrates on finding satellite galaxies that belong to the same dark matter
halo (see \citealt{Yang:2007p506}), it could return groupings that are mainly
gravitationally bound. Detection of extended X-ray radiation can also indicate gravitationally bound groups.

\begin{acknowledgements}
SMN acknowledges the funding by Finland's Academy of Sciences and Letters and
the Nordic Optical Telescope (NOT) Scientific Association (NOTSA). SMN would like
to thank Dr. Gerard Lemson for invaluable help with the Millennium Simulation
database. SMN would also like to acknowledge the support and help provided by Dr.
Pekka Hein\"am\"aki and Dr. Pasi Nurmi. We thank the
anonymous referee for detailed reading of the manuscript and comments that
helped us to improve the original manuscirpt. The Millennium Simulation
databases used in this paper and the web application providing online access 
were constructed as part of the activities of the German Astrophysical Virtual Observatory.
\end{acknowledgements}

\bibliographystyle{aa} 
\bibliography{paper2} 

\end{document}